\documentstyle[12pt,subeq,epsfig]{article}

\begin{document}

\title{Thermodynamic Properties of the Two-Dimensional Coulomb Gas
in the Low-Density Limit}

\author{
P. Kalinay$^1$ and L. {\v S}amaj$^{1,2}$
}

\maketitle

\begin{abstract}
The model under consideration is the two-dimensional Coulomb gas 
of $\pm$ charged hard disks with diameter $\sigma$.
For the case of pointlike charges $(\sigma=0)$, the system is 
stable against collapse of positive-negative pairs of charges in 
the range of inverse temperatures $0 \le \beta < 2$, where its 
full thermodynamics was obtained exactly 
[L. {\v S}amaj and I. Trav{\v e}nec, {\it J. Stat. Phys.}
{\bf 101}:713 (2000)].
In the present work, we derive the leading correction to
the exact thermodynamics of pointlike charges due to the
presence of the hard core $\sigma$ (appearing in the
dimensionless combination $n\sigma^2$, $n$ is the particle density).
This permits us to extend the treatment to the interval
$2\le \beta <3$ (the Kosterlitz-Thouless phase transition
takes place at $\beta=4$).
The results, which are exact in the low-density limit 
$n\sigma^2 \to 0$, reproduce correctly the singularities 
of thermodynamic quantities at the collapse point $\beta=2$ 
and agree very well with Monte-Carlo simulations.
\end{abstract}

\medskip

\noindent {\bf KEY WORDS:} Coulomb gas; thermodynamics;
charge pairing; low-density limit, sum rule.

\vfill

\noindent $^1$ Institute of Physics, Slovak Academy of Sciences, 
D\'ubravsk\'a cesta 9, 842 28 Bratislava, Slovakia;
E-mails: (P.K.) fyzipali@savba.sk, (L.{\v S}.) fyzimaes@savba.sk

\noindent $^2$ Laboratoire de Physique Th\'eorique, Universit\'e
de Paris-Sud, B\^atiment 210, 91405 Orsay Cedex, France

\newpage

\renewcommand{\theequation}{1.\arabic{equation}}
\setcounter{equation}{0}

\section{Introduction and strategy}
The model under consideration is the two-dimensional Coulomb gas
(2dCG), i.e., a neutral system of positive and negative unit charges
$q_i = \pm 1$ in a plane, interacting through the pair potential
\begin{equation} \label{1.1}
v({\bf r}_i,{\bf r}_j) = \left\{ 
\begin{array} {ll}
-q_i q_j \ln \left( \vert {\bf r}_i - {\bf r}_j \vert / L \right) , & 
\vert {\bf r}_i - {\bf r}_j \vert > \sigma \cr
\infty , &  \vert {\bf r}_i - {\bf r}_j \vert \le \sigma
\end{array} \right.
\end{equation} 
Here, the logarithmic Coulomb potential is the solution of
the 2d Poisson equation $\Delta v({\bf r}) = - 2 \pi \delta({\bf r})$.
The Coulomb potential is regularized at short distance by 
a hard-core potential of diameter $\sigma$ around each charge.
The system is studied as the classical one and in thermodynamic
equilibrium, via the grand canonical ensemble characterized by
the (dimensionless) inverse temperature $\beta$ and the couple
of equal particle fugacities $z_+ = z_- = z$.
The fugacity $z$ has dimension [length]$^{-2}$.
Within the grand canonical formalism, the length scale $L$ in 
(\ref{1.1}) manifests itself as the rescaling of $z$,
$z \to L^{\beta/2} z$.
We shall set $L$ to unity for simplicity, keeping in mind
that the true dimension of the rescaled $z$ is 
[length]$^{\beta/2-2}$. 
The corresponding particle number densities $n_+ = n_- = n/2$
($n$ is the total particle density) enter into the formalism
in the dimensionless combination $n\sigma^2$.

For small values of the dimensionless
density $n\sigma^2$, the famous Kosterlitz-Thouless
(KT) transition \cite{Kosterlitz} of infinite order takes place 
at a specific density-dependent inverse temperature 
$\beta_{\rm KT}$ \cite{Levin,Alastuey}.
In the high-temperature conducting phase $\beta < \beta_{\rm KT}$,
the effective potential between infinitesimal external charges
decays exponentially due to the perfect screening by the positive and
negative charges of the Coulomb system.
In the low-temperature dielectric phase $\beta > \beta_{\rm KT}$,
the system charges form dipoles and no longer screen an external charge, 
so that the effective potential between infinitesimal external charges
is proportional to the bare logarithmic potential.
At high enough $n\sigma^2$, the KT critical
line splits into a first order liquid-gas coexistence curve
(for Monte-Carlo (MC) simulations, see refs. \cite{Caillol,Orkoulas},
for theoretical computation, see ref. \cite{Levin}).

In the low-density limit $n\sigma^2 \to 0$,
which is of special interest in general and also in this paper,
the thermodynamic behavior of the 2dCG as a function of $\beta$
undergoes fundamental changes at two points: $\beta_{\rm c} = 2$
(the collapse of pointlike particles) and $\beta_{\rm KT} = 4$
(the KT phase transition).
The first -- collapse -- point reflects the fact that, for the
case of strictly pointlike particles $\sigma = 0$, the
singularity of the Coulomb potential $v({\bf r})$ (\ref{1.1})
at the origin prevents the thermodynamic stability against
collapse of positive-negative pairs of charges (or, equivalently,
the corresponding Boltzmann factor $r^{-\beta}$ is not integrable
at short distances in 2d) for $\beta \ge 2$.
Thus, for $0\le \beta<2$, the system of pointlike particles is 
thermodynamically stable and the introduction of a hard core around 
particles is a marginal perturbation which does not change the 
thermodynamics substantially.
On the other hand, for $2\le \beta <4$, the introduction of a hard core
is inevitable for avoiding the collapse:
when one calculates thermodynamic quantities and at the end takes
the limit $\sigma\to 0$ (with $z$ being fixed), while the density,
the free energy and the internal energy per particle diverge
due to the collapse phenomenon, the specific heat and the truncated 
(Ursell) correlation functions are expected to remain finite \cite{Hauge}.
In spite of the tendency to the collapse into neutral pairs of charges, 
there still exist free charges which are able to screen
and the system remains in its conducting phase up to the KT phase 
transition at point $\beta_{\rm KT}=4$.
We would like to stress that, for a given fugacity $z$ and
when $\beta \ge 2$, although $n\to \infty$ as $\sigma\to 0$,
the dimensionless density goes to the limit of interest,
$n\sigma^2 \to 0$.

In what follows, we shall summarize in detail the known results
in the two qualitatively different regimes $(0\le \beta<2)$ and 
$(2< \beta <4)$, and at the collapse point $\beta_c=2$.

In the stability range of inverse temperatures $0\le \beta<2$,
as has been already mentioned, the thermodynamics of the 2dCG
is well defined even for the case of pointlike particles,
$\sigma=0$.
The density derivatives of the Helmholtz free energy, like the
pressure $p$, can be calculated exactly by using a simple
scaling argument.
For instance, the equation of state
\begin{equation} \label{1.2}
\beta p = n \left( 1 - {\beta\over 4} \right)
\end{equation}
has been known for a very long time \cite{Salzberg}.
The temperature derivatives of the Helmholtz free energy,
like the internal energy $U$ or the constant volume 
(surface in 2d) specific heat $C_V$, are nontrivial quantities,
the calculation of which can be based on an explicit 
density-fugacity relationship.
This relationship was obtained only recently \cite{Samaj1}
via a mapping of the 2dCG onto a classical 2d sine-Gordon
theory with a specific normalization of the cos-field, and
then by using quite recent results about that integrable
field theory \cite{Destri,Zamolodchikov1}.
Explicitly,
\begin{eqnarray} \label{1.3}
{n \over z^{4/(4-\beta)}} & = & 
\left( {\pi \beta \over 8} \right)^{\beta/(4-\beta)} 
\left[ 2 ~ {\Gamma(1-\beta/4) \over \Gamma(1+\beta/4)} 
\right]^{4/(4-\beta)} \nonumber \\
& & \times {\Gamma^2\left( 1+\beta/[2(4-\beta)]\right)
\over (1/\pi) \Gamma^2\left( 1/2+\beta/[2(4-\beta)]\right)}
{{\rm tg}\left( \pi\beta/[2(4-\beta)]\right) \over
\pi\beta/[2(4-\beta)]}
\end{eqnarray}
where $\Gamma$ stands for the Gamma function
The density-fugacity relationship (\ref{1.3}) was checked
on a few lower orders of its high-temperature $\beta$-expansion
by using a renormalized Mayer expansion in {\it density}, 
valid just in the stability regime.
For fixed $z$, the particle density exhibits the expected
collapse singularity as $\beta \to 2^-$:
\begin{equation} \label{1.4}
n \sim {4\pi z^2 \over 2-\beta}
\end{equation}
This behavior can be derived by using an independent-pair picture 
of the system around the collapse point \cite{Hauge},
which is another check of the exact results.
Based on the density-fugacity relationship (\ref{1.3}),
the complete thermodynamics of the pointlike CG can be obtained by
elementary means in the whole stability interval
$0 \le \beta < 2$ \cite{Samaj1}.

At the collapse point $\beta_c=2$, by the continualization of
Gaudin's lattice model \cite{Gaudin}, which is expected
to have the same properties as the 2dCG in the low-density limit,
the truncated many-body densities (Ursell functions) were found
in refs. \cite{Cornu1,Cornu2}.
These densities have the remarkable property of going to 
well-defined limits as $n\sigma^2$ vanishes (as is believed,
this property lasts up to the KT phase transition),
identical to the densities of an equivalent Thirring model
at the free-fermion point.
The knowledge of all truncated many-body densities at $\beta_c=2$
for $n\sigma^2=0$ permits one to extract the leading parts
of thermodynamic quantities at $\beta=2$ which, for a
fixed fugacity $z$, do not vanish in the low-density limit 
$n\sigma^2\to 0$ \cite{Cornu1}.
Namely,
\begin{subequations} \label{1.5}
\begin{eqnarray}
n & = & 4 \pi z^2 \left[ \ln \left( {1\over \sigma \pi z} \right)
- C + O(1) \right] \label{1.5a} \\
\beta p & = & 2 \pi z^2 \left[ \ln \left( {1\over \sigma \pi z} \right)
- C + {1\over 2} + O(1) \right] \label{1.5b} \\
u^{\rm ex} & = & {1\over 4} \left[ \ln \left( {\sigma \over \pi z} \right)
- C + O(1) \right] \label{1.5c} \\
{c_V^{\rm ex}\over k_B} & = & {1\over 6} 
\left[ \ln \left( {1\over \sigma \pi z} \right) - C \right]^2
- {1\over 4} \left[ \ln \left( {1\over \sigma \pi z} \right) - C \right]
- {1\over 8} + O\left( {1\over \ln(\sigma\pi z)}\right) \label{1.5d}
\end{eqnarray}
\end{subequations}
Here, $u^{\rm ex} = \langle E \rangle/N$ is the excess (over ideal)
internal energy per particle, $c_V^{\rm ex} = C_V^{\rm ex}/N$
is the excess specific heat at constant volume per particle, and
$C$ is the Euler constant.

The region of inverse temperatures $2\le \beta <4$ is 
usually studied by using the Mayer series expansion of the specific
grand potential in {\it fugacity}.
It was proven that each term of the $z$-series converges
in the insulator region $\beta>4$ \cite{Gallavotti,Speer}.
For $\beta\le 4$, the existence of infinitely many thresholds at
inverse temperatures
\begin{equation} \label{1.6}
\beta_l = 4 \left( 1 - {1\over 2l} \right),
\quad \quad l = 1, 2, \ldots
\end{equation}
lying between $\beta_1 = \beta_c = 2$ and 
$\beta_{\infty} = \beta_{\rm KT} = 4$, was observed: 
if $\beta>\beta_l$, only the Mayer series' coefficients (cluster
integrals) up to the order $2l$ are finite, and the cluster integrals 
of order $> 2l$ exhibit the large-distance divergence 
in the infinite-volume limit.
Below the collapse point $\beta<2$ (where the density 
format is appropriate) all cluster integrals diverge.
The free energy is supposed to have at points 
$\{ \beta_l \}_{l=1}^{\infty}$ a logarithmic dependence on the
cut-off (in our case $n\sigma^2$) \cite{Justin}.
Points $\{ \beta_l \}_{l=1}^{\infty}$ were conjectured to correspond
to a sequence of transitions from the pure multipole insulating
phase $(\beta>4)$ to the conducting phase $(\beta<2)$ via an infinite
number of intermediate phases \cite{Gallavotti}.

Such a conjecture was later denied by Fisher et al. \cite{Fisher}.
In the limit $n\sigma^2 \to 0$, they proposed the following
ansatz for the equation of state in the fugacity 
format:\footnote{we rescale $z$ by $\sigma^{(4-\beta)/2}$ 
in the ansatz in order to work with dimensionless quantities}
\begin{eqnarray} \label{1.7}
\beta p & = & b_{\psi}(\beta) z^{2\psi(\beta)}
\left[ 1 + e(z\sigma^{(4-\beta)/2},\beta) \right] \nonumber \\
& & + {1\over \sigma^2} \sum_{l=1}^{\infty} {\bar b}_{2l}(\beta)
\left[ z \sigma^{(4-\beta)/2} \right]^{2l}
\end{eqnarray}
where the temperature-dependent exponent is
\begin{equation} \label{1.8}
2 \psi(\beta) = {4\over 4-\beta}
\end{equation}
The function $e$ vanishes as $n\sigma^2 \to 0$.
At points $\beta_l$ [see relation (\ref{1.6})],
$\psi(\beta_l) = l$.
The coefficient $b_{\psi}(\beta)$ and the function
$e(z\sigma^{(4-\beta)/2},\beta)$ were suggested to be analytic 
in the whole conducting regime $0\le \beta <4$.
The coefficients $\{ {\bar b}_{2l}(\beta)\}_{l=1}^{\infty}$
were suggested to be zero for $0\le \beta<2$, and finite,
analytic in $\beta$, for $2\le \beta <4$.
The divergence of the coefficients of an {\it analytic}
expansion in $z^2$ was related to the appearance of
the {\it anomalous} term $b_{\psi} z^{2\psi}$ in (\ref{1.7}).
Moreover, at $\beta_{\rm KT}=4$, $\psi(\beta_{KT})$ becomes
infinite and the anomalous term disappears from the ansatz,
in agreement with the results of refs. \cite{Gallavotti,Speer}.
The findings of this work support the ansatz (\ref{1.7}),
however, the supposed analytic behavior of the coefficients
$b_{\psi}(\beta)$ and $\{ {\bar b}_{2l}(\beta)\}_{l=1}^{\infty}$
in the whole interval of $\beta \in \langle 0,4)$ is apparently wrong.
For example, in the stability region $0\le \beta <2$ and
in the limit of pointlike particles $n\sigma^2 = 0$,
the equation of state (\ref{1.2}), when combined with the
ansatz (\ref{1.7}) with $e=0$, yields
\begin{equation} \label{1.9}
b_{\psi}(\beta) = \left( 1 - {\beta\over 4} \right)
{n \over z^{4/(4-\beta)}}
\end{equation}
Substituting $n/z^{4/(4-\beta)}$ by the rhs of (\ref{1.3}),
we get the exact 
$b_{\psi}(\beta)\propto {\rm tg}(\pi\beta/[2(4-\beta)])$.
This function has the simple pole just at the collapse point
$\beta_1 = \beta_c = 2$, and its analytic continuation to the
region $2<\beta<4$ has the simple poles at $\{ \beta_l \}_{l=2}^{\infty}$
given by (\ref{1.6}) (see also the next paragraph).

The aim of this paper is to derive the leading correction to 
the exact thermodynamics of pointlike charges due to the presence
of the hard core of diameter $\sigma$, and in this way to extend
the treatment of the 2dCG beyond the collapse point $\beta_c=2$.
First of all, we have to understand the analytic structure of
a ``naive'' continuation of the exact results for the particle
density of pointlike charges (\ref{1.3}) to the range of inverse 
temperatures $2\le \beta <4$, in order to see which singularities 
should be removed by the consideration of a hard core.
The only source of singularities in $n$ (\ref{1.3}) is
the tg-function whose decomposition into simple fractions
\cite{Gradshteyn} gives
\begin{equation} \label{1.10}
{{\rm tg}(\pi\beta/[2(4-\beta)]) \over \pi\beta/[2(4-\beta)]}
= {2\over \pi^2} \sum_{l=1}^{\infty} {1\over (\beta_l-\beta)}
~ {(4-\beta)^2 (4-\beta_l)^2 \over \beta(4-\beta_l) +
\beta_l (4-\beta)}
\end{equation} 
The thresholds $\{ \beta_l \}_{l=1}^{\infty}$, given by
(\ref{1.6}), manifest themselves as simple poles at which
$n$ exhibits the discontinuity from
$\lim_{\beta\to \beta_l^-} n(\beta) \to \infty$ to
$\lim_{\beta\to \beta_l^+} n(\beta) \to -\infty$.
The introduction of the hard core $\sigma$ must remove the artificial 
singularities of $n$ at $\{ \beta_l \}_{l=1}^{\infty}$ 
and must make $n$ positive as is required by its basic definition.

We reach our aim by combining the zeroth-moment (electroneutrality)
sum rule for the pair charge-charge correlation function
\cite{Stillinger}, valid in the whole fluid regime 
$0\le \beta \le 4$, with the short-distance expansion of that
charge-charge correlation.
The short-distance expansion of the charge-charge correlation
can be done, in principle, systematically.
Every new term of order ``$l$'' should remove the singularities of 
the pointlike $n$ at $\beta_l$ and makes the particle density
positive up to $\beta_{l+1}$.
We were able to establish the leading term of order ``1'' which
permits us to extend the treatment of the 2dCG to the interval
$2\le \beta <3$.
The advantage of our approach with respect to the standard ones
based on the fugacity expansion is that, in the whole interval
$0\le \beta <3$, our estimates are exact in the low-density
limit $n\sigma^2 \to 0$.
Moreover, the leading parts of thermodynamic quantities
at $\beta=2$ (\ref{1.7}) are reproduced correctly.
The agreement with the MC simulations \cite{Caillol} is very good. 

The paper is organized as follows.
The method is described in Section 2.
The complete thermodynamics is derived and then tested in various
limits in Section 3.
In section 4, the comparison of our results is made with the MC 
simulations and some concluding remarks are given.

\renewcommand{\theequation}{2.\arabic{equation}}
\setcounter{equation}{0}

\section{Method}
In the conducting phase of the two-component plasma with
the Coulomb plus an arbitrary short-range pair interaction of
particles, the zeroth and second moments of the
charge-charge density are determined exclusively by the long-range
tail of the Coulomb potential \cite{Stillinger}.
The starting point of our calculation is the zeroth-moment
(electroneutrality) sum rule
\begin{equation} \label{2.1}
n_q = \int \left[ U_{q,-q}({\bf r}) - U_{q,q}({\bf r}) \right]
{\rm d}^2 r
\end{equation}
Here, with the notation 
${\hat n}_q({\bf r}) = \sum_i \delta_{q,q_i} \delta({\bf r}-{\bf r}_i)$ 
for the microscopic density of particles of charge $q=\pm 1$ at
position ${\bf r}$, $n_q = \langle {\hat n}_q({\bf r}) \rangle = n/2$
and the Ursell function 
$U_{q,q'}({\bf r},{\bf r}') = U_{q,q'}(\vert {\bf r}-{\bf r}' \vert)$
(denoted as $\rho_{q,q'}^{(2){\rm T}}$ in refs. \cite{Cornu1,Cornu2})
is defined by
\begin{equation} \label{2.2}
U_{q,q'}({\bf r},{\bf r}') = 
\langle {\hat n}_q({\bf r}) {\hat n}_{q'}({\bf r}') \rangle
- n_q \delta_{q,q'} \delta({\bf r}-{\bf r}') - n_q n_{q'} 
\end{equation}
For the Coulomb system of interest (\ref{1.1}), the difference
$U_{q,-q}-U_{q,q}$ vanishes inside the hard core, and the
total particle number density $n$ is given by
\begin{equation} \label{2.3}
n(z,\sigma) = 2 \int_{\sigma}^{\infty} 2 \pi r {\rm d}r
\left[ U_{q,-q}(r;z,\sigma) - U_{q,q}(r;z,\sigma) \right]
\end{equation}
Hereinafter, we omit in the notation the dependence of
quantities on $\beta$.

The dependence of the density $n$ on $\sigma$ in (\ref{2.3}) comes
from the cutoff in the integration over $r$ and from the 
$\sigma$-dependence of the Ursell functions themselves.
These Ursell functions are supposed to be well defined and finite 
for the zero density, $n\sigma^2 = 0$, 
in the whole conducting regime $0\le \beta <4$, including the 
collapse interval $2\le \beta <4$.
This belief is strongly supported by the finite values
of the Ursell functions at the collapse point $\beta_c=2$
\cite{Cornu1,Cornu2}:
\begin{subequations} \label{2.4}
\begin{eqnarray}
U_{q,-q}(r;z,0) & = & \left( {m^2\over 2\pi} \right)^2 
K_1^2(m r)  \label{2.4a} \\
U_{q,q}(r;z,0) & = & - \left( {m^2\over 2\pi} \right)^2 
K_0^2(m r)  \label{2.4b}
\end{eqnarray}
\end{subequations}
with fixed $m = 2 \pi z$, $K_0$ and $K_1$ are modified Bessel functions.
We can thus write
\begin{equation} \label{2.5}
U_{q,q'}(r;z,\sigma) = U_{q,q'}(r;z,0) + 
\Delta_{q,q'}(r;z,\sigma) , \quad \quad
r\ge \sigma
\end{equation}
which defines $\Delta_{q,q'}(r;z,\sigma)$, vanishing when
$\sigma\to 0$, as the change of $U_{q,q'}(r;z,0)$ due to
the introduction of the hard core $\sigma\le r$ to pointlike
particles.
Subtracting Eq. (\ref{2.3}) with $\sigma> 0$ and the same
equation with $\sigma=0$, one arrives at
\begin{eqnarray} \label{2.6}
n(z,\sigma) - n(z,0) & = & - 2 \int_0^{\sigma} 2\pi r {\rm d}r
\left[ U_{q,-q}(r;z,0) - U_{q,q}(r;z,0) \right] \nonumber \\
& & + 2 \int_{\sigma}^{\infty} 2 \pi r {\rm d}r
\left[ \Delta_{q,-q}(r;z,\sigma) - \Delta_{q,q}(r;z,\sigma) \right]
\end{eqnarray}
Since $n(z,0)$ is defined only in the stability regime
$0\le \beta <2$, for the time being we shall restrict ourselves
to this range of $\beta$.
We now make a heuristic assumption analogous to that made
at the point $\beta_c=2$ in refs. \cite{Cornu1,Cornu2}:
in the low-density limit $n\sigma^2\to 0$ and for
$0\le \beta< 4$, one can neglect the quantities
$\Delta_{q,\pm q}$ in Eq. (\ref{2.6}).
Consequently,
\begin{equation} \label{2.7}
n(z,\sigma) = n(z,0) - 2 \int_0^{\sigma} 2\pi r {\rm d}r
\left[ U_{q,-q}(r;z,0) - U_{q,q}(r;z,0) \right] 
\end{equation}
Our assumption is equivalent to saying that in equation (\ref{2.6}) 
only the contribution $\propto \int_0^{\sigma} r {\rm d}r
[U_{q,-q}(r;z,0) - U_{q,q}(r;z,0)]$ with $\sigma>0$ is enough for 
removing, via a systematic short-distance expansions of
$U_{q,\pm q}(r;z,0)$, term by term the singularities
of $n(z,0)$ at points $\{ \beta_l \}_{l=1}^{\infty}$
(see the Introduction).
This scenario will be verified at the first singular point
$\beta_1 = \beta_c = 2$.

The next step is to construct the short-distance expansions
of $U_{q,\pm q}(r;z,0)$ in the integral on the rhs of (\ref{2.7}).
For small enough $\beta$, the short-distance expansion of
the Ursell functions is dominated by the Boltzmann factor of
the corresponding pair Coulomb potential \cite{Jancovici,Hansen}.
In the case of oppositely charged particles, one has
\begin{equation} \label{2.8}
U_{q,-q}(r;z,0) \sim z^2 r^{-\beta} \quad \quad
{\rm as}\ r\to 0
\end{equation}
valid in the whole interval $0\le \beta <4$.
Note that since $K_1(mr) \sim 1/(m r)$ for $r\to 0$,
$U_{q,-q}(r;z,0)$ at $\beta_c=2$ [see relation (\ref{2.4a})]
satisfies (\ref{2.8}).
In the case of the same charges, the leading term has
a more complicated structure,
\begin{equation} \label{2.9}
U_{q,q}(r;z,0) \propto \left\{
\begin{array}{ll}
r^{\beta} & {\rm for}\ 0\le \beta <1 \cr
r^{2-\beta} & {\rm for}\ 1\le \beta <2
\end{array}
\quad \quad {\rm as}\ r\to 0 \right.
\end{equation}
The change of the power-law behavior is caused by the divergence of 
the prefactor to $r^{\beta}$ at $\beta=1$ \cite{Hansen}.
Considering in (\ref{2.4b}) that $K_0(m r) \sim - \ln r$
for $r\to 0$, the logarithmic behavior of $U_{q,q}(r;z,0)$
at $\beta_c=2$ can be understood as a limiting case of
$r^{2-\beta}$ in (\ref{2.9}).
Within the sine-Gordon representation of the 2d pointlike CG,
the formula corresponding to (\ref{2.8}) is known as the
conformal normalization of the cos-field.
For such a theory, the short-distance expansion of correlation
functions is available by using the Operator Product Expansion
\cite{Wilson}, as was explicitly done in ref. \cite{Fateev}.
Although this method allows one to construct systematically
the short-distance expansions of $U_{q,\pm q}(r;z,0)$ with
coefficients expressed in terms of Dotsenko-Fateev integrals
\cite{Dotsenko}, it applies only to small values of $\beta$
and does not describe, e.g., the change of the behavior at
$\beta=1$ (\ref{2.9}).
In any case, in the interval $0\le \beta <4$, the charge-charge
combination of the Ursell functions is dominated at short
distances by (\ref{2.8}),
\begin{equation} \label{2.10}
U_{q,-q}(r;z,0) - U_{q,q}(r;z,0) \sim z^2 r^{-\beta}
\quad \quad {\rm as}\ r\to 0
\end{equation}
and we shall consider just this leading term.

Inserting (\ref{2.10}) into (\ref{2.7}), one obtains the
basic formula
\begin{equation} \label{2.11}
n(z,\sigma) = n(z,0) - 4\pi z^2 {\sigma^{2-\beta} \over 2-\beta}
\end{equation}
Although this result was derived in the region $0\le \beta<2$
where $n(z,0)$ is well defined by (\ref{1.3}), it is reasonable
to assume the continuation of (\ref{2.11}) beyond the collapse point 
$\beta_c=2$: both the sum rule (\ref{2.3}) and the leading 
short-distance expansion (\ref{2.10}), which play the crucial role 
in the derivation of (\ref{2.11}), remain valid up to the KT transition
at $\beta_{\rm KT}=4$.
As will be shown in the subsequent section, the leading correction
term in Eq. (\ref{2.11}) removes the singularity of $n(z,0)$
at $\beta_1 = \beta_c = 2$, and provides an adequate description
of the 2dCG with $n\sigma^2$ small up to $\beta_2=3$ where
another singularity of $n(z,0)$ occurs.
This singularity at $\beta_2=3$ should be removed by the next
term of the short-distance expansion (\ref{2.10}) inserted into
(\ref{2.7}), however, as was mentioned above, it is not simple
to get the explicit form of that term for such a large value
of $\beta$.

\renewcommand{\theequation}{3.\arabic{equation}}
\setcounter{equation}{0}

\section{Thermodynamics}
Based on the density-fugacity relationship (\ref{2.11}),
we derive in this section the thermodynamics of the 2dCG
in the restricted region of interest $0\le \beta <3$.
Our findings will be compared with the known results and
conjectures reviewed in the Introduction, namely with
the exact formulae at $\beta_c=2$ (\ref{1.5}), with the
predictions based on the independent-pair collapse picture
for $\beta>2$ and in the limit $\sigma\to 0$ \cite{Hauge},
and with the proposal (\ref{1.7}) for the equation of state.
The stability regime $0\le \beta <2$ will not be discussed
in detail since there the introduction of the hard core to
pointlike particles is a marginal perturbation.

We first express relation (\ref{1.3}) for the density of
pointlike particles $n(z,0)$ in a more convenient form,
\begin{equation} \label{3.1}
n(z,0) = {4\pi \Phi(\beta) \over 2-\beta} z^{4/(4-\beta)}
\end{equation}
where we have introduced the function
\begin{eqnarray} \label{3.2}
\Phi(\beta) & = & 
\left( {\pi \beta \over 8} \right)^{\beta/(4-\beta)} 
\left[ 2 ~ {\Gamma(1-\beta/4) \over \Gamma(1+\beta/4)} 
\right]^{4/(4-\beta)} {2-\beta \over 4\pi} \nonumber \\
& & \times {\Gamma^2\left( 1+\beta/[2(4-\beta)]\right)
\over (1/\pi) \Gamma^2\left( 1/2+\beta/[2(4-\beta)]\right)}
{{\rm tg}\left( \pi\beta/[2(4-\beta)]\right) \over
\pi\beta/[2(4-\beta)]}
\end{eqnarray}
With regard to the collapse singularity (\ref{1.4}), $\Phi$
was choosen such that $\Phi(\beta=2)=1$.
The Taylor expansion of $\ln \Phi(\beta)$ around $\beta=2$
thus reads
\begin{eqnarray} \label{3.3}
\ln \Phi(\beta) & = & (\ln \pi + C) (\beta-2) + 
{1\over 2} (\ln \pi + C) (\beta-2)^2 \nonumber \\
& & + {1\over 4} \left[ \ln \pi + C - {17\over 12} \zeta(3) \right]
(\beta-2)^3 + \cdots
\end{eqnarray}
The function $\Phi(\beta)$ is positive in the interval
$0\le \beta < 8/3$, it crosses zero at point $\beta=8/3$
(which is not exceptional from any point of view) 
and diverges to $-\infty$ as $\beta\to 3$.
Using the representation (\ref{3.1}) in Eq. (\ref{2.11}),
one gets for $n=n(z,\sigma)$
\begin{equation} \label{3.4}
n = {4\pi \over 2-\beta} z^{4/(4-\beta)}
\left[ \Phi(\beta) - \left( \sigma z^{2/(4-\beta)}
\right)^{2-\beta} \right]
\end{equation}
Notice that this relation is dimensionally correct --
it is expressible in terms of dimensionless quantities
\begin{equation} \label{3.5}
\xi = \sigma z^{2/(4-\beta)}
\end{equation}
and $n\sigma^2$, or the packing fraction
\begin{equation} \label{3.6}
\eta = {1\over 4} \pi n \sigma^2
\end{equation}
used in the MC simulations \cite{Caillol}.
Taking into account the expansion (\ref{3.3}) in Eq. (\ref{3.4})
one sees that, for fixed $z$ and nonzero $\sigma$, the density
$n$ is finite at $\beta_c=2$; its value coincides with the
expected result (\ref{1.5a}).
Formula (\ref{3.4}) can be analytically continued to the
region $2<\beta<3$.
For $\beta>2$ and small $\sigma$, the term 
$[\sigma z^{2/(4-\beta)}]^{2-\beta}$ becomes larger than
$\Phi(\beta)$ and, when combined with the negative
denominator $(2-\beta)$, it implies the positive sign of $n$
up to $\beta=3$.
When $\sigma\to 0$, this term and consequently $n$ diverge
in the region $2\le \beta<3$ as a consequence of the collapse
phenomenon.

We proceed by the derivation of the equation of state.
The grand canonical potential $\Omega$ is determined by
the thermodynamic relation
\begin{equation} \label{3.7}
n = z {\partial (-\beta \Omega/V) \over \partial z}
\end{equation}
where $V (\to \infty)$ is the volume (in 2d, the surface)
of the homogeneous system.
With respect to the boundary condition
$(-\beta\Omega/V)\vert_{z=0} = 0$, the integration of (\ref{3.7}),
with $n$ substituted from Eq. (\ref{3.4}), results in the
equation of state for $\beta p = -\beta\Omega/V$,
\begin{equation} \label{3.8}
\beta p = {\pi (4-\beta) \over 2-\beta} \Phi(\beta) z^{4/(4-\beta)}
- {2\pi \over \sigma^2 (2-\beta)} 
\left( z \sigma^{(4-\beta)/2} \right)^2
\end{equation}
Eq. (\ref{3.8}) has the form of the ansatz (\ref{1.7}) with
\begin{equation} \label{3.9}
b_{\psi} = {\pi (4-\beta) \over 2-\beta} \Phi(\beta),
\quad \quad {\bar b}_2 = - {2\pi \over 2-\beta}
\end{equation}
In contrast to the conjecture made in ref. \cite{Fisher},
these coefficients are not analytic functions of $\beta$ 
at $\beta=2$.
Let us now rewrite the equation of state (\ref{3.8}) with the aid
of the density-fugacity relationship (\ref{3.4}) as follows
\begin{equation} \label{3.10}
{\beta p \over n} = {1\over 2} + 
{(2-\beta) \Phi \over 4(\Phi - \xi^{2-\beta})}
\end{equation}
where $\xi$ is defined by (\ref{3.5}).
At $\beta_c=2$ and for nonzero $\sigma$, (\ref{3.10}) agrees
with relations (\ref{1.5a}) and (\ref{1.5b}).
In the limit of pointlike particles $\sigma\to 0$, our equation
of state (\ref{3.10}) reduces to the one obtained by Hauge
and Hemmer \cite{Hauge}: in the stability regime $0\le \beta<2$,
the hard-core correction $\xi^{2-\beta}$ is negligible for small
$\sigma$ and the $\sigma\to 0$ limit corresponds to (\ref{1.2}),
while for $2< \beta <3$ $\xi^{2-\beta}$ diverges when
$\sigma\to 0$ and so $\beta p/n \to 1/2$, i.e., the pairs
of $\pm 1$ charged particles collapse into neutral ``free''
particles of half density $n/2$.

Finally, the (excess) dimensionless specific free energy
$f = \beta F^{\rm ex}/N$ reads
\begin{eqnarray} \label{3.11}
f(n,\beta) & = & {\beta \Omega \over n V} + \ln z \nonumber \\
& = & - {1\over 2} - {(2-\beta) \Phi \over 4(\Phi - \xi^{2-\beta})}
+ \ln z 
\end{eqnarray} 
where the implicit dependence of the fugacity $z(n,\beta)$
on the particle density $n$ is determined by Eq. (\ref{3.4}).
According to the elementary thermodynamics, the (excess)
internal energy per particle $u^{\rm ex}$ is given by
$u^{\rm ex} = \partial f(n,\beta) / \partial \beta$, explicitly
\begin{equation} \label{3.12}
u^{\rm ex} = {1\over 2} \ln \sigma - {1\over 2(2-\beta)}
- {1\over 2(\Phi-\xi^{2-\beta})} \left[ \left( 2-{\beta\over 2}
\right) \Phi' + \Phi \ln \xi \right]
\end{equation} 
and the (excess) specific heat at constant volume per particle
$c_V^{\rm ex}$ is given by $c_V^{\rm ex}/k_B = -\beta^2
\partial^2 f(n,\beta) / \partial \beta^2$, explicitly
\begin{eqnarray} \label{3.13}
{1\over \beta^2} {c_V^{\rm ex}\over k_B} & = &
{1\over 2 (2-\beta)^2} + {(4-\beta)\Phi''\over 4 (\Phi-\xi^{2-\beta})}
\nonumber \\
& & - {1\over 2 (2-\beta) (\Phi - \xi^{2-\beta}) 
[ \Phi - (2-\beta/2)\xi^{2-\beta} ] } \nonumber \\
& & \times \Big\{ (2-\beta) \Phi' \left[ \Phi + 
\left( 2 - {\beta\over 2} \right) \Phi' \right] +
(\Phi - \xi^{2-\beta}) {\Phi\over 2} \nonumber \\
& & \quad + (2-\beta) \xi^{2-\beta} (\ln \xi)
\left[ (1+\ln \xi) \Phi + (4-\beta) \Phi' \right] \Big\} 
\end{eqnarray}
At $\beta_c=2$ and for nonzero $\sigma$, $u^{\rm ex}$ is identical
to (\ref{1.5c}), while
\begin{equation} \label{3.14}
{c_V^{\rm ex}(\beta=2) \over k_B} = {1\over 3}
{ [\ln(\pi z\sigma) + C]^3 + 2 [\ln(\pi z\sigma)+C]^2
\over 1 + 2 [\ln(\pi z\sigma)+C]}
- {17\over 12} \zeta(3) {1\over [\ln(\pi z\sigma)+C]}
\end{equation}
The expansion of this expression into the Laurent
series in $1/[\ln(\pi z \sigma)+C]$ reproduces the leading
terms in (\ref{1.5d}).
In the region $2<\beta<3$ and in the limit $\sigma\to 0$,
$u^{\rm ex}$ diverges to $-\infty$ due to the term
$(\ln \sigma)/2$ involving the energy of the collapsed pair
of particles.
$c_V^{\rm ex}$ behaves differently: $\xi^{2-\beta}$ becomes
large for small $\sigma$ and kills all terms on the rhs of 
(\ref{3.13}), except of the first one.
The consequent finite result 
$c_V^{\rm ex}/k_B = \beta^2/[2(2-\beta)^2]$
coincides with the finding of ref. \cite{Hauge}.

\renewcommand{\theequation}{4.\arabic{equation}}
\setcounter{equation}{0}

\section{Comparison with MC simulations and conclusion}
In the previous two sections, we have derived the thermodynamics of
the 2dCG by adding the leading hard-core correction term to
the exact density-fugacity relationship for the Coulomb system
of pointlike particles.
We have successfully tested various limits of our results.
In this section, the comparison is made with the MC simulations
of the 2dCG \cite{Caillol}.

\begin{figure}
\epsfig{file=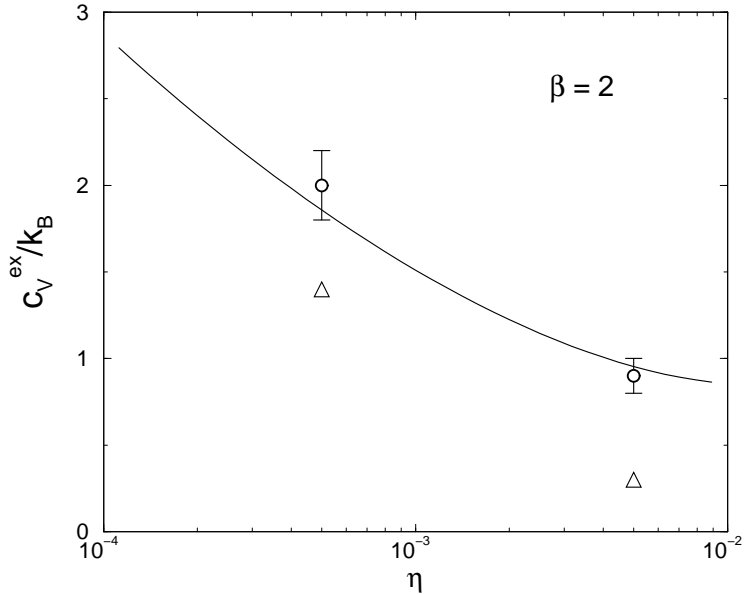, width=10cm}
\caption{The plot of $c_V^{\rm ex}/k_B$ vs. the packing
fraction $\eta$ at $\beta_c=2$: present result (\ref{3.14})
(solid line), formula (\ref{1.5d}) (triangles) and MC simulations
(circles).}
\end{figure}
The results for the heat capacity $c_V^{\rm ex}$ at the collapse
point $\beta_c=2$, as the function of the packing fraction $\eta$
(\ref{3.6}), are presented in Fig. 1.
Our formula (\ref{3.14}), represented by the solid line, provides
the estimates of $c_V^{\rm ex}$ consistent with the MC data
(circles) within the error bars.
The agreement with the MC simulations is much better than in
the case of the previously derived formula (\ref{1.5d})
(triangles), which involves the first three terms of the
Laurent expansion of (\ref{3.14}).

\begin{figure}
\epsfig{file=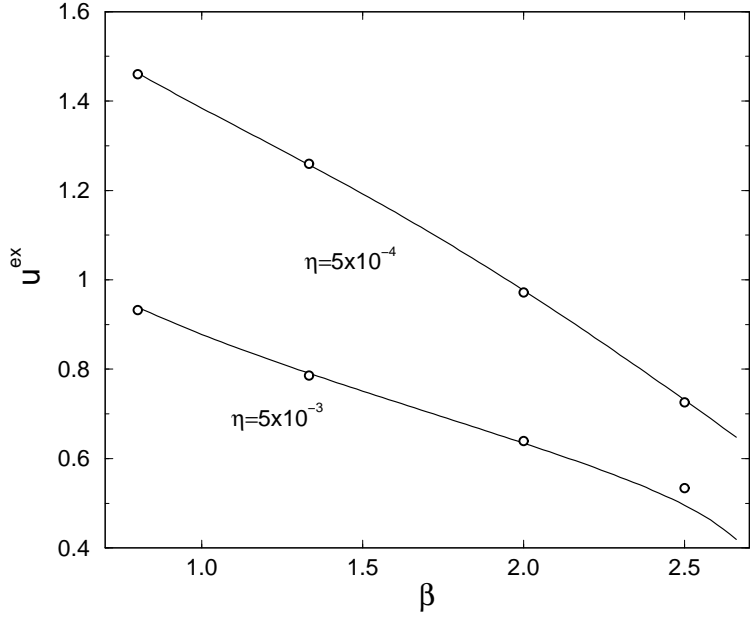, width=10cm}
\caption{The plot of $u^{\rm ex}$ vs. the inverse temperature
$\beta$ for $\eta=5\times 10^{-4}$ and $5\times 10^{-3}$:
present result (\ref{3.12}) (solid lines) and MC simulations
(circles).}
\end{figure}
\begin{figure}
\epsfig{file=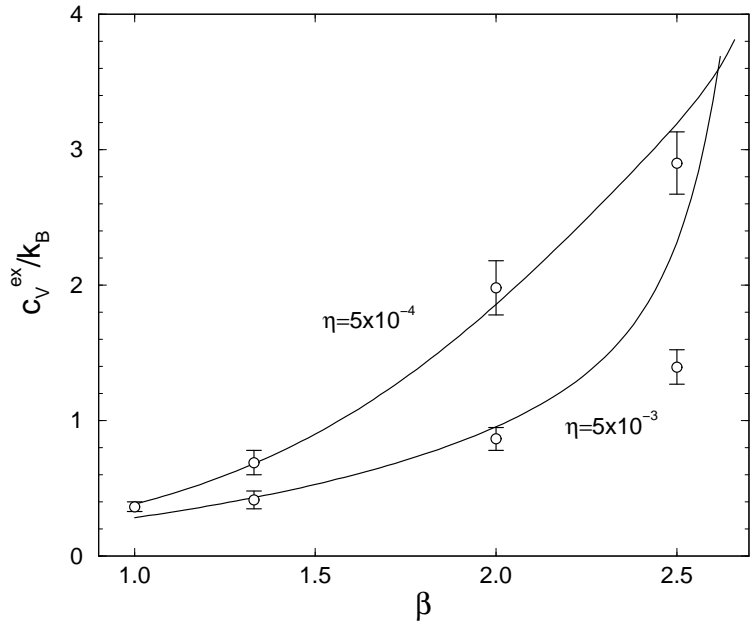, width=10cm}
\caption{The plot of $c_V^{\rm ex}/k_B$ vs. the inverse temperature
$\beta$ for $\eta=5\times 10^{-4}$ and $5\times 10^{-3}$:
present result (\ref{3.13}) (solid lines) and MC simulations
(circles).}
\end{figure}
In Figs. 2 and 3, we represent by solid lines our plots of the
internal energy $u^{\rm ex}$ (\ref{3.12}) and the heat capacity
$c_V^{\rm ex}$ (\ref{3.13}), respectively, versus the inverse
temperature $\beta$, for two values of the packing fraction
$\eta = 5\times 10^{-4}$ and $5\times 10^{-3}$.
The agreement with the MC data (circles) is very good, 
but getting worse when approaching point $\beta=3$:
here, the next (neglected) term of the hard-core corrections,
which eliminates the next singularity of $n(z,0)$ at
$\beta_2=3$, starts to be important.

In conclusion, the extension of the thermodynamic treatment
of the 2dCG around and beyond $\beta_2=3$, up to $\beta_{\rm KT}=4$,
requires to construct a systematic short-distance expansion of 
the Ursell functions for pointlike charges.
It is clear that the relevant hard-core corrections must
subtract all simple poles of the pointlike-particle
density $n(z,0)$ (\ref{1.3}) at $\{ \beta_l \}_{l=1}^{\infty}$
given by (\ref{1.6}).
Thus, the density-fugacity relationship should be of the form
\begin{equation} \label{4.1}
n(z,\sigma) = n(z,0) - \sum_{l=1}^{\infty}
{{\bar c}_{2l}(\beta) \over \beta_l-\beta}
\left( z^2 \sigma^{\beta_l-\beta} \right)^l
\end{equation}
where the coefficient ${\bar c}_{2l}(\beta)$ $(l=1, 2,\ldots)$
is regular at $\beta=\beta_l$.
According to our result (\ref{2.11}), ${\bar c}_2(\beta)=4\pi$.
Taking into account the formula (\ref{1.3}) for $n(z,0)$,
the exponent of $z$, $4/(4-\beta)$, is equal to $2 l$ at
the pole $\beta=\beta_l$ and the exponent of $\sigma$ is
put by dimensional reasons.
Note that since $n = z \partial (\beta p)/\partial z$,
(\ref{4.1}) automatically leads to the ansatz (\ref{1.7})
with $e=0$ [because we have not taken into account in (\ref{2.6}) 
``slight'' hard-core changes of the correlation functions 
$\Delta_{q,\pm q}$ defined by (\ref{2.5})]. 
The coefficients $b_{\psi}$ and $\{ {\bar b}_{2l} \}_{l=1}^{\infty}$
are {\it singular} functions of $\beta$.
The $l$th term in the sum on the rhs of (\ref{4.1}) has
the correct behavior in the limit $\sigma\to 0$:
it goes to 0 when $\beta<\beta_l$ and removes the singularity
of $n(z,0)$ at $\beta=\beta_l$ (and simultaneously gives rise to
a logarithmic dependence on the hard core at this point).
A systematic generation of the coefficients ${\bar c}_l(\beta)$
is our task for the future. 

\section*{Acknowledgments}
We are very indebted to B. Jancovici for valuable discussions,
careful reading of the manuscript and useful comments.
We thank also A. Alastuey and P. J. Forrester for stimulating
discussions.
The stay of L. {\v S}amaj in LPT Orsay was supported by a 
NATO fellowship.
The support by Grant VEGA 2/7174/20 is acknowledged.

\end{document}